\documentclass[aps,pre,reprint,onecolumn,nofootinbib,notitlepage,showpacs,showkeywords]{revtex4-1}
\usepackage{hyperref}
\usepackage{amsfonts}
\usepackage{amssymb}
\usepackage{amsmath}
\usepackage{graphicx}
\usepackage{epsfig}
\usepackage{tensor}
\usepackage{color}
\usepackage[T1]{fontenc}
\usepackage[latin9]{inputenc}

\usepackage{xparse}

\NewDocumentCommand{\tens}{t_}
 {%
  \IfBooleanTF{#1}
   {\tensop}
   {\otimes}%
 }
\NewDocumentCommand{\Log}{o}{%
  \IfNoValueTF{#1}{}{{}^{#1}\!}\log}%

\newcommand{\ket}[1]{| #1 \rangle}

\begin{document}

\title{Multiple transitions between normal and hyperballistic diffusion in quantum walks with time-dependent jumps}
\author{Marcelo A. Pires}
\author{S\'{i}lvio M. \surname{Duarte~Queir\'{o}s} }
\thanks{Associate to the National Institute of Science and Technology for Complex Systems, Brazil}
\affiliation{Centro Brasileiro de Pesquisas F\'{\i}sicas \\
		Rua Dr Xavier Sigaud, 150, 22290-180 Rio de Janeiro --- RJ, Brazil}
\author{Giuseppe Di Molfetta}
\affiliation{Aix-Marseille Universit\'e, Universit\'e de Toulon, CNRS,
LIS, Marseille, France, Natural Computation research group}

\begin{abstract}
We extend to the gamut of functional forms of the probability distribution of the time-dependent step-length a previous model dubbed {\it Elephant Quantum Walk},
which considers a uniform distribution and yields hyperballistic dynamics where the variance grows cubicly with time, $\sigma ^2 \propto t^3$, and a Gaussian for the position of the walker.
We investigate this proposal both locally and globally with the results showing that the time-dependent interplay between interference, memory and long-range hopping leads to multiple transitions between dynamical regimes, namely ballistic $\rightarrow$ diffusive $\rightarrow$ superdiffusive $\rightarrow$  ballistic $\rightarrow$ hyperballistic for  non-hermitian coin whereas the first diffusive regime is quelled for implementations using the Hadamard coin.
In addition, we observe a robust asymptotic approach to maximal coin-space entanglement. 
\end{abstract}


\date{ \today}

\maketitle

\section{\label{sec:level1}Introduction}

In his seminal article on quantum computing, Richard Feynman \cite{feynman1982simulating} suggested computers which use quantum logic for information processing may be employed to simulate quantum systems efficiently, even when that is impossible to computers based on classical logic. To simulate the dynamics of a quantum system usually means to describe the system in terms of qubits --- as well as its dynamics --- by a succession of local and unitary operations, involving at most two qubits at time. In recent decades, Quantum Walks (QWs) --- or its multi-particle generalization ---, namely quantum cellular automata, have become the most natural way to design a wide range of complex phenomena and extensively used for their comprehension. QWs are frequently translated into simple models which act as proxies for rather complex dynamics as those governed by quantum fields. Besides offering easily implementable physical protocols, such approach has opened new avenues for the fundamental understanding of those processes. For this reason, research in QWs has bridged disciplines such as natural calculus and algorithmics \cite{montanaro2016quantum,santha2008quantum,childs2009universal,ambainis2007quantum}, quantum field theory \cite{di2013quantum,di2016quantum,marquez2017fermion,marquez2018electromagnetic} and discrete geometry \cite{bru2016quantum, arrighi2018dirac,Arrighi2018quantumwalkingin}, complex systems \cite{faccin2013degree,di2015nonlinear,caruso2014universally} and machine learning \cite{paparo2014quantum,belovs2014power}.
Formally, a QW describes the unitary dynamics of one quantum particle and its internal degrees of freedom. The key content of a discretisation unit --- or 'cell' --- is whether or not the particle occupies that cell and what its internal state is. Moreover, as for any quantum system, these properties may be found in superposition. In a single time step the particle can only move a finite distance \cite{grossing1988quantum, meyer1996quantum, aharonov1993quantum}. That protocol was then systematically extended on graphs \cite{aharonov2001quantum} and later fully mathematically examined by \cite{konno2002quantum}. 

A significant interpretation of QWs concerns looking at them as mathematical frameworks to study dissipative quantum computing/simulation \cite{sinayskiy2012efficiency, kendon2003decoherence}. Explicitly, in most of the cases which are worth studying --- namely those related to small/nano-systems ---, the noise level is high. Thus, if purely coherent quantum computing is a long-standing goal, the modelling and simulation of efficiently noisy quantum systems is a mid-term objective and thus a relevant milestone in that research path. Few, yet important, results have been obtained in this direction, mainly focussing on one particle models, which although rich, may still have inherent limits; for instance, the authors in Ref.~\cite{attal2012open} have generalised the usual noisy unitary QW (e.g. \cite{di2016discrete}) to the so-called Open QWs, in order to describe the environment interaction and \cite{uchiyama2018environmental} has recently understood to what extent the environmental noise can enhance quantum transport in photosynthetic complexes, i.e., how mimicking natural open quantum systems (mostly biological systems) can optimise quantum information processing.     

Complementary, it has surged an interest for super and hyper ballistic phenomena in some kind of physical systems, such as viscous electron flows and in some kind of disordered and quasiperiodic system \cite{guo2017higher}.

Within this same context, some of the authors (GDM and SMDQ) introduced \cite{di2018elephant} an analytically treatable non-Markovian discrete-time QW in a one-dimensional lattice which yields hyper-ballistic diffusion that is characterized by a variance growing cubicly with time, $\sigma^2 \propto t^3$. The key ingredient in that model is a temporal noise uniformly distributed in the translation operator. For its rules, our model can be understood as the quantum version of the classical non-Markovian 'elephant random walk' process \cite{schutz2004elephants}.

In the following, we extend 
this class of Elephant 
Quantum Walk (EQW) to a more general family of the 
functional noise, namely the discretised 
$q$-exponential, so that a 
quite broad range of 
asymptotic behaviour can be 
analyzed. For brevity, we call it generalized EQW or simply 
gEQW. That functional form 
bounded by both the 
delta-Dirac distribution --- 
associated with the standard 
QW --- and the 
uniform distribution, which 
yields the EQW.

\section{\label{sec:model}Model}

We consider a QW over the $(1+1)$--spacetime grid. Its coin or spin degree of freedom lies $\mathcal{H}_2$, for which we may chose some orthonormal basis $\{\ket{v^L}, \ket{v^R}\}$. The overall state of the walker lies the composite Hilbert space $\mathcal{H}_2\otimes \mathcal{H}_\mathbb{Z}$ and may be thus be written $\Psi=\sum_x \psi^+(x) \ket{v_L}\otimes\ket{x} + \psi^-(x) \ket{v_R}\otimes\ket{x}$, where the scalar field $\psi^L$ (resp. $\psi^R$) gives, at every position $x\in \mathbb{Z}$, the amplitude of the particle being there and about to move left (resp. right). We use $(t,x) \in \mathbb{N} \times \mathbb{Z}$, to label respectively instants and points in space and let:

\begin{equation}
\Psi_{t+1}=W_{t'} \Psi_t
\label{eq:FDE0}
\end{equation}
where 
\begin{equation}
W_{t'} = \hat{S}(\hat{C}\otimes\text{Id}_\mathbb{Z})
\end{equation}
with
$\hat{S}$ a state-dependent shift operator such that 
\begin{equation}
(\hat{S}_{t^\prime}\Psi)_{t}(x)=\begin{pmatrix}\psi^L_t(x-t^\prime)\\\psi^R_t(x+t^\prime)\end{pmatrix}.
\end{equation} 
where $t'$ represents the walker displacement on the lattice.

In the following we will consider two family of coin operators in $U(2)$:

\begin{equation}
\widehat C_{H}
= 
\begin{pmatrix}
\cos \theta &   \sin \theta \\
 \sin \theta  & -\cos \theta
\end{pmatrix}
\quad\quad
\widehat C_{K}
= 
\begin{pmatrix}
\cos \theta &  i \sin \theta \\
i \sin \theta  & \cos \theta
\end{pmatrix} .
\end{equation}

respectively, the Hadamard-like coin $\widehat C_{K}$ and a non-hermitian coins family $\widehat C_{K}$ already introduced e.g. in \cite{kempe2003quantum}. Notice that in the following we choose the following localised initial state:
\begin{equation}
 \Psi_{0}(x) = \frac{1}{\sqrt{2}} \delta_{x,0}  \left( |0\rangle + e^{i\phi} |1\rangle  \right) \tens{} |x\rangle ,  
 \label{Eqt0}
 \end{equation}
 and, in order to have symmetric distributions, we will set $\phi=\pi/2$ for $\widehat C_{H} $ and $\phi=0$ for $\widehat C_{K}$.

In the family of EQWs introduced in \cite{di2018elephant} the parameter $t^\prime$ is the memory parameter, which at every interval $[1,t]$ follows
the functional form
\begin{equation}
\mathcal{P}_{[1,t]}(k) \equiv
\mathcal{C}_t \exp_q(-k)  = \mathcal{C}_t \left[ 1- (1-q)\,k \right]_{+}^{1/(1-q)},  \quad \text{with} \quad  k=\{1,2,\ldots, t \},
\label{Eq:kernel_qexp}
\end{equation}
known as a $q$-exponential distribution as well. The quantity $\mathcal{C}_t$ is a time-dependent normalisation that must be updated at each iteration. 
Furthermore, from Eq.\ref{Eq:kernel_qexp} we observe that as $\min(k)=1 \rightarrow \min(t^\prime)=1$ then the minimum allowed step has length 1.
Interpreting it as a kernel distribution, Eq.~(\ref{Eq:kernel_qexp}) has been applied in econometric~\cite{queiros2007generalised} as well as in cellular automata models~\cite{rohlf2007long}. 

Looking at the functional form Eq.~(\ref{Eq:kernel_qexp}) we can identify  the following traits: for $q<1$, it has a 'compact' support (in the sense that for $t \gg 1$ the maximum value given by Eq.~(\ref{Eq:kernel_qexp}) is less than $t$),, where the maximum value one can select is equal to the integer of $\left(\frac{1}{1-q}\right)$; for $q = 1$, it gets an exponential form and is natural scale-dependence. For those two instances, all the statistical moments are finite (when $t \rightarrow \infty$). For $t>1$, Eq.~(\ref{Eq:kernel_qexp}) exhibits an asymptotic power-law decay for which the $n$-th order statistical moment is not finite for $ q \ge \frac{2+n}{1+n} $.
That said, we understand that for $q=1/2$ we will have the same dynamics as the standard QW as we can only consider the nearest neighbour whereas the limit $q \rightarrow + \infty$ concurs with the EQW case.

\section{\label{sec:results}Results and discussion}

\subsection{Global and local properties}

\begin{figure}[t]
\includegraphics[width=0.59\linewidth]{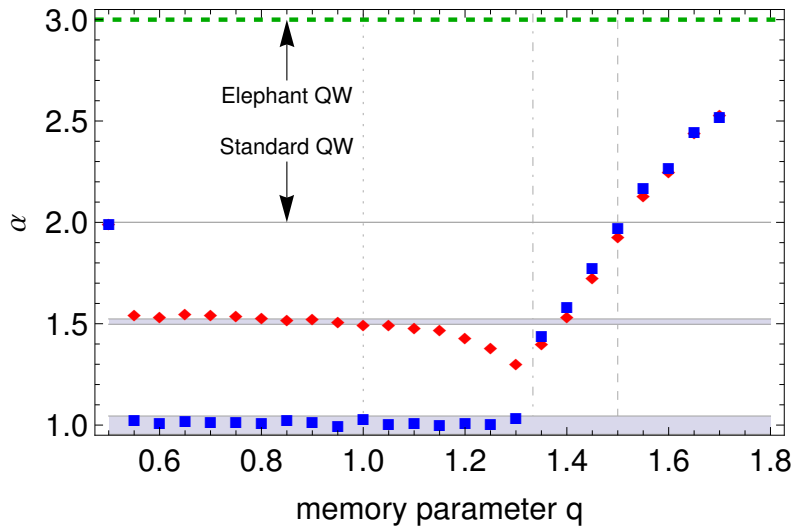}
\caption{Dynamic regimes for $ \widehat C_{H} $ (red) and $ \widehat C_{K} $ (blue) coins. We use $t_{max}=10^4$ to estimate $\alpha$ from $\overline{x^2}\sim t^\alpha$.  Interestingly, jumps play a dual role in the QW: inhibition  of wavepacket spreading for short-range steps and and enhancement of spreading for long-range hopping.}
\label{fig:diagram-alpha}
\end{figure}

\begin{figure}[htbp]
\centering
\includegraphics[width=0.71\linewidth]{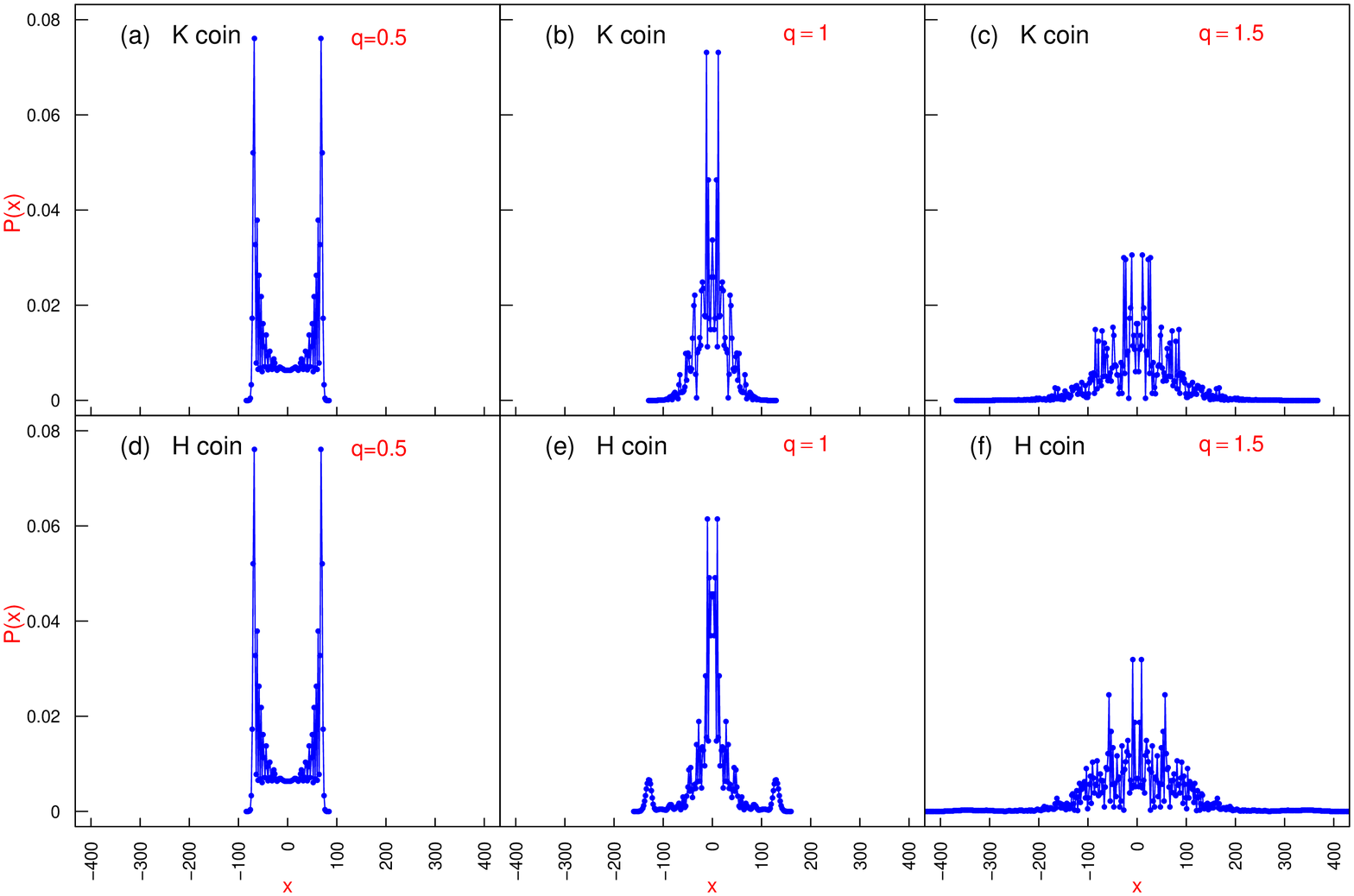}
\caption{Probability distribution $P_t(x)$  at $t=100$. Panels show typical profiles for  coins $ \widehat C_{H} $ and $ \widehat C_{K} $.}
\label{fig:pxt}

\medskip

\includegraphics[width=0.71\linewidth]{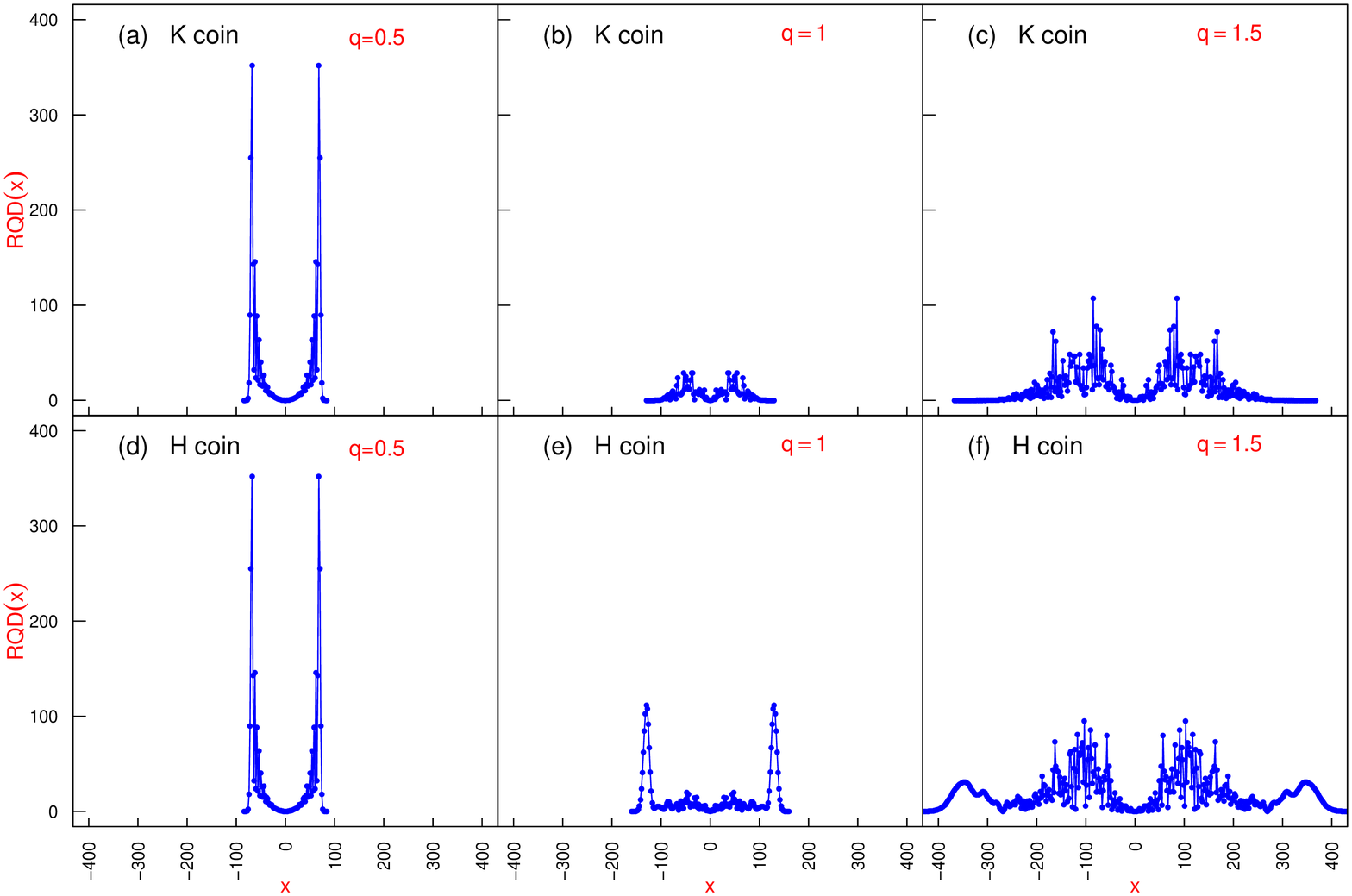}
\caption{Local relative quadratic deviation $RQD(x)$  at $t=100$.  Panels show typical profiles for the corresponding $P_t(x)$ in Fig \ref{fig:pxt}. }
\label{fig:rwdx}
\end{figure}

\begin{figure}[htbp]
\centering

\includegraphics[width=0.81\linewidth]{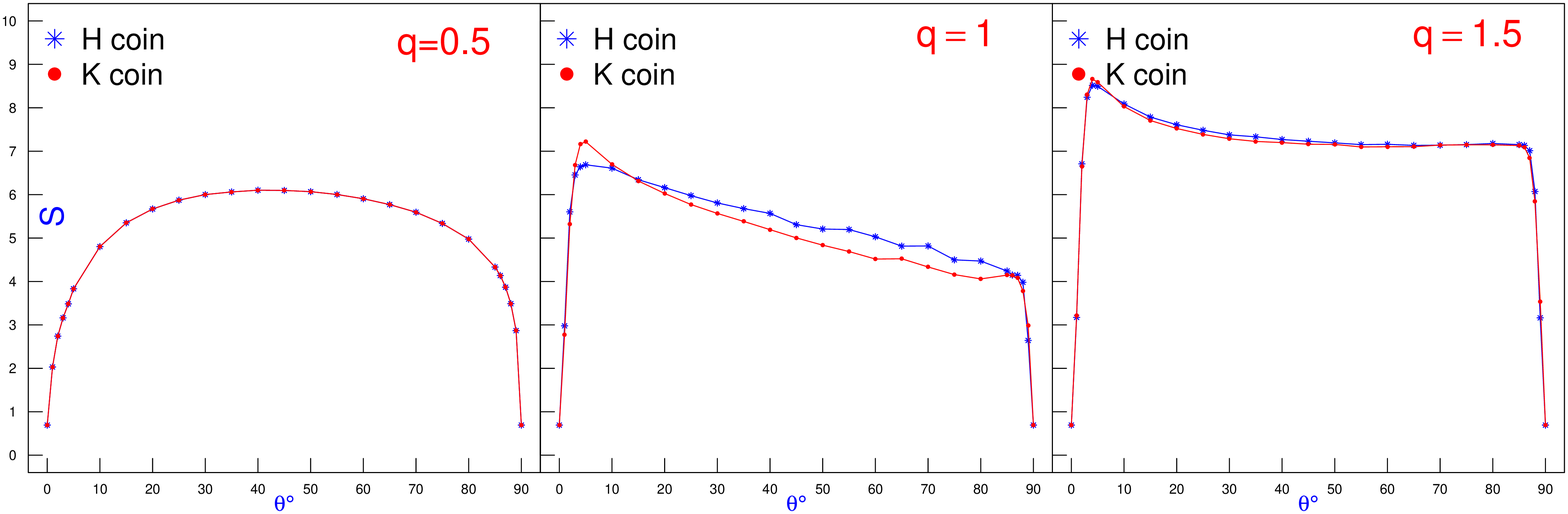}
\caption{Shannon entropy $S$ versus the coin angle $\theta$  for $ \widehat C_{H} $ and $ \widehat C_{K} $ and $q=\{0.5,1,1.5\}$. For the memoryless case $q=0.5$ we recover the same behaviour for $S$ as shown  in Fig.5 of \cite{chandrashekar2008optimizing}.}
\label{fig:Shannon}

\medskip

\includegraphics[width=0.81\linewidth]{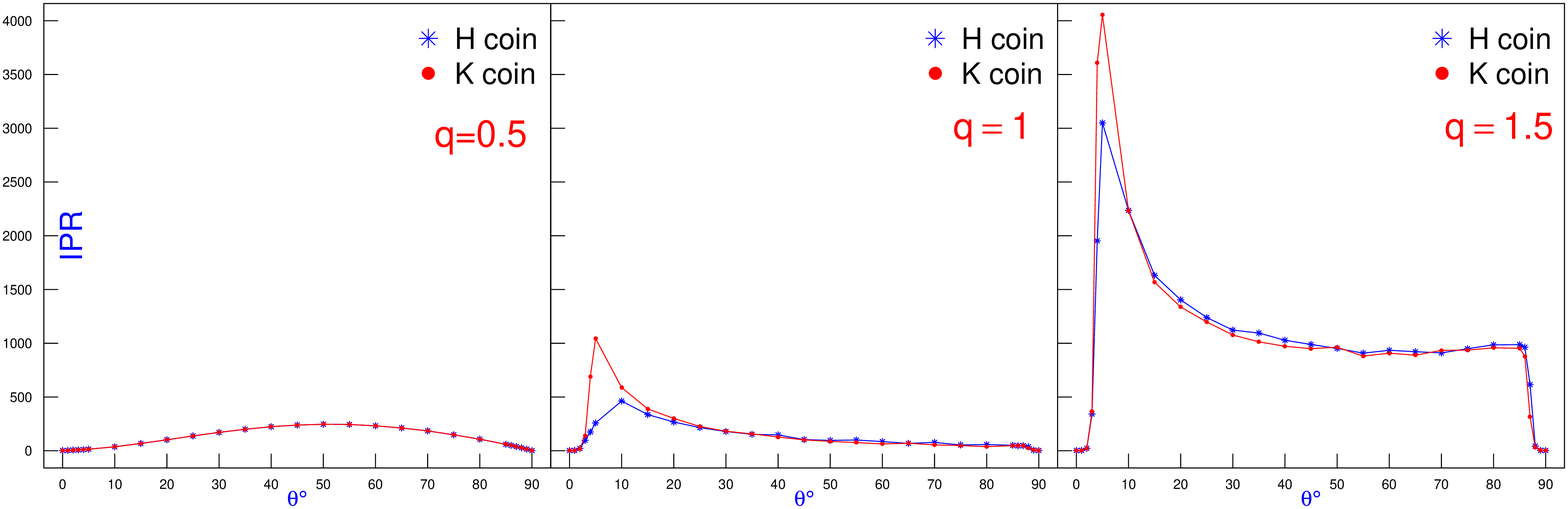}
\caption{Inverse participation ratio $IPR$ versus the coin angle $\theta$. Panels show results for $ \widehat C_{H} $ and $ \widehat C_{K} $ and $q=\{0.5,1,1.5\}$.}
\label{fig:IPR}

\medskip

\includegraphics[width=0.81\linewidth]{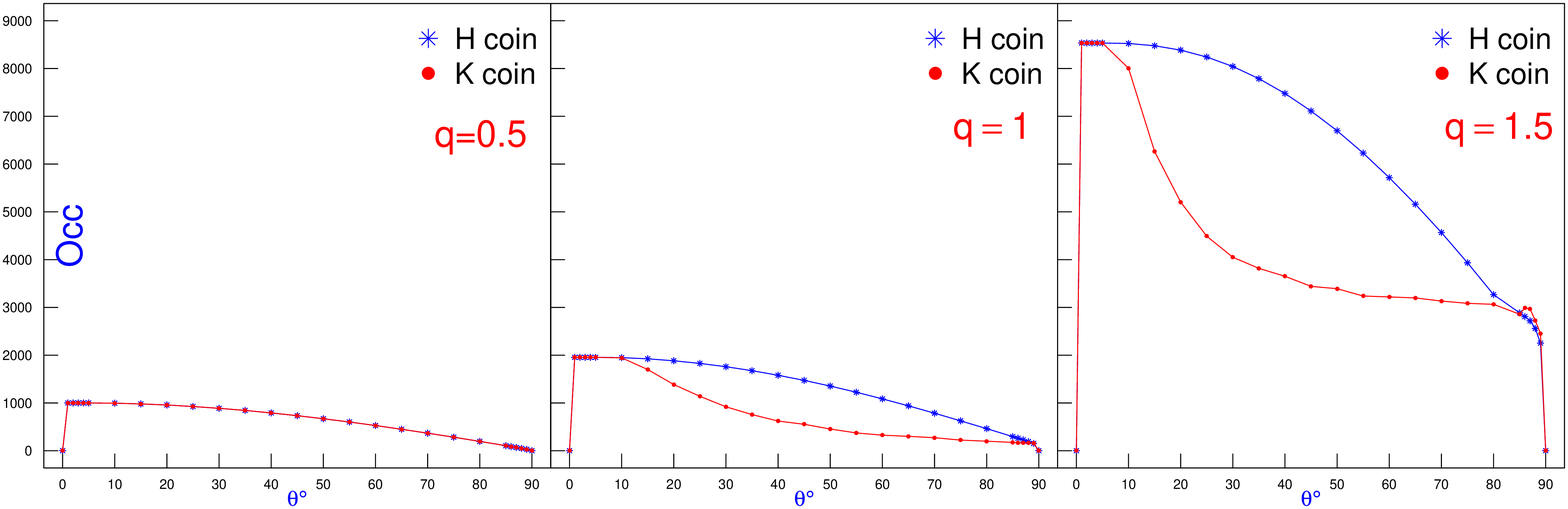}
\caption{Occupancy $Occ$ versus the coin angle $\theta$. Panels show results for $ \widehat C_{H} $ and $ \widehat C_{K} $ and $q=\{0.5,1,1.5\}$.}
\label{fig:occupancy}

\end{figure}


\begin{figure}[t]
\includegraphics[scale=0.35]{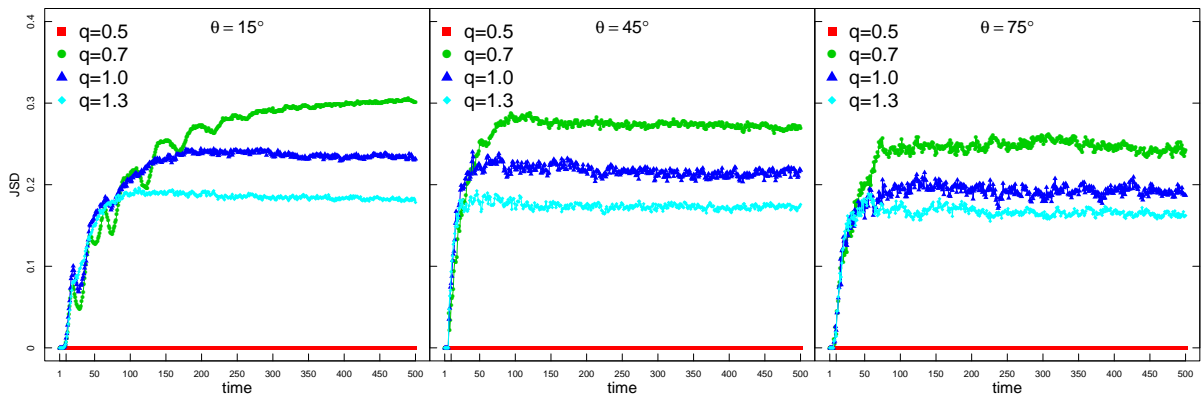}
\caption{Time series for the Jensen-Shannon Dissimilarity (JSD) between  $P_t^{H}(x)$  and $P_t^{K}(x)$ for typical angles  $\theta=\{15^o,45^o,75^o\}$ and $q=\{0.5,0.7,1,1.3\}$.}
\label{fig:jsd}
\end{figure}

\begin{figure}[b]
\centering
\includegraphics[scale=0.45]{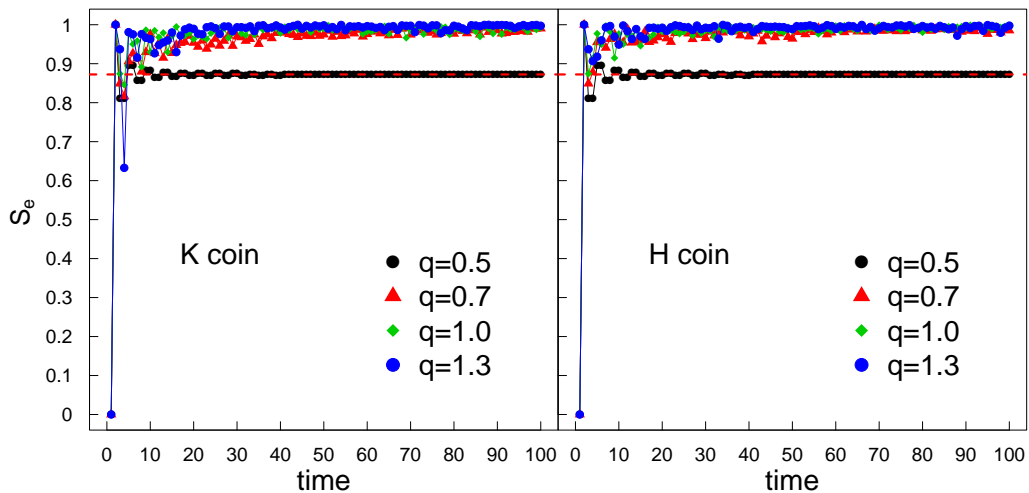}
\caption{Time series for the Von Neumann entanglement entropy $S_e$ obtained from the reduced density  matrix. The   horizontal dashed red line shows  the analytic value for the usual  QW shown in Eq.23 of \cite{abal2006quantum}.}
\label{fig:entang}
\end{figure}

\begin{figure}[b]
\centering
\includegraphics[scale=0.45]{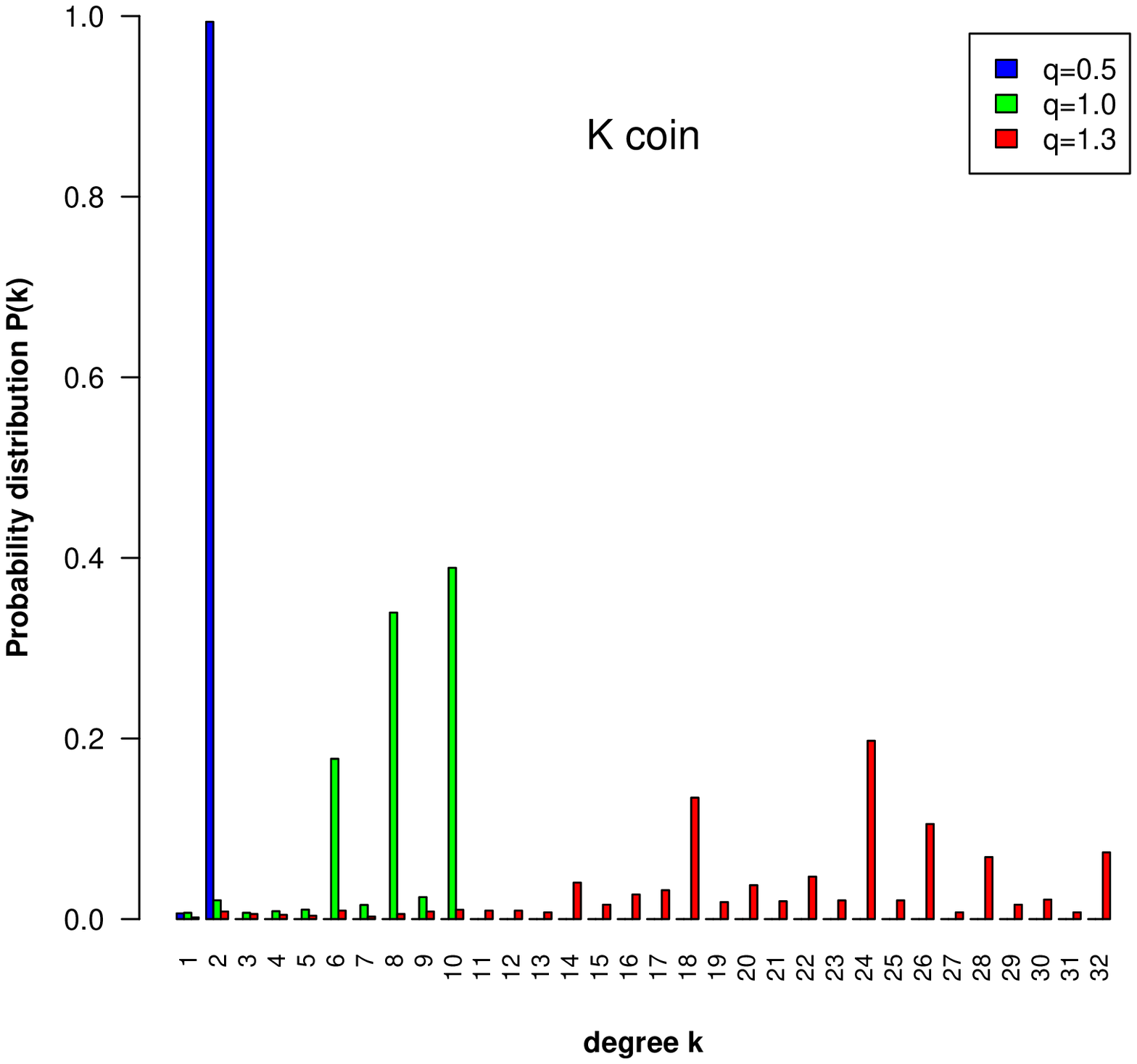}
\includegraphics[scale=0.45]{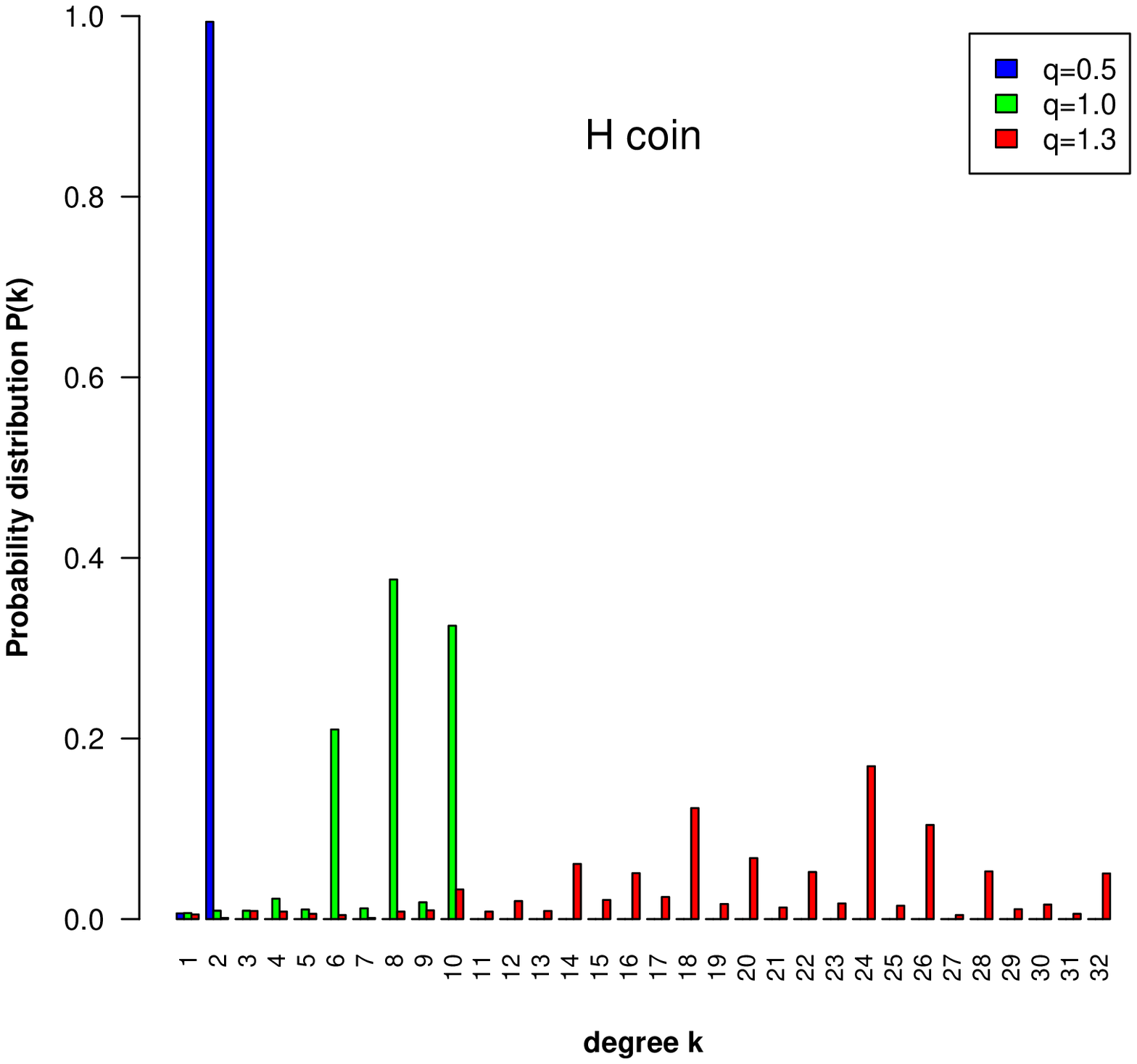}
\caption{Degree distribution at $t=200$ of the networks constructed from the QW with jumps  for $ \widehat C_{H} $ and $ \widehat C_{K} $.}
\label{fig:net-degree}
\end{figure}

\begin{figure}[t]
\centering
\includegraphics[scale=0.48]{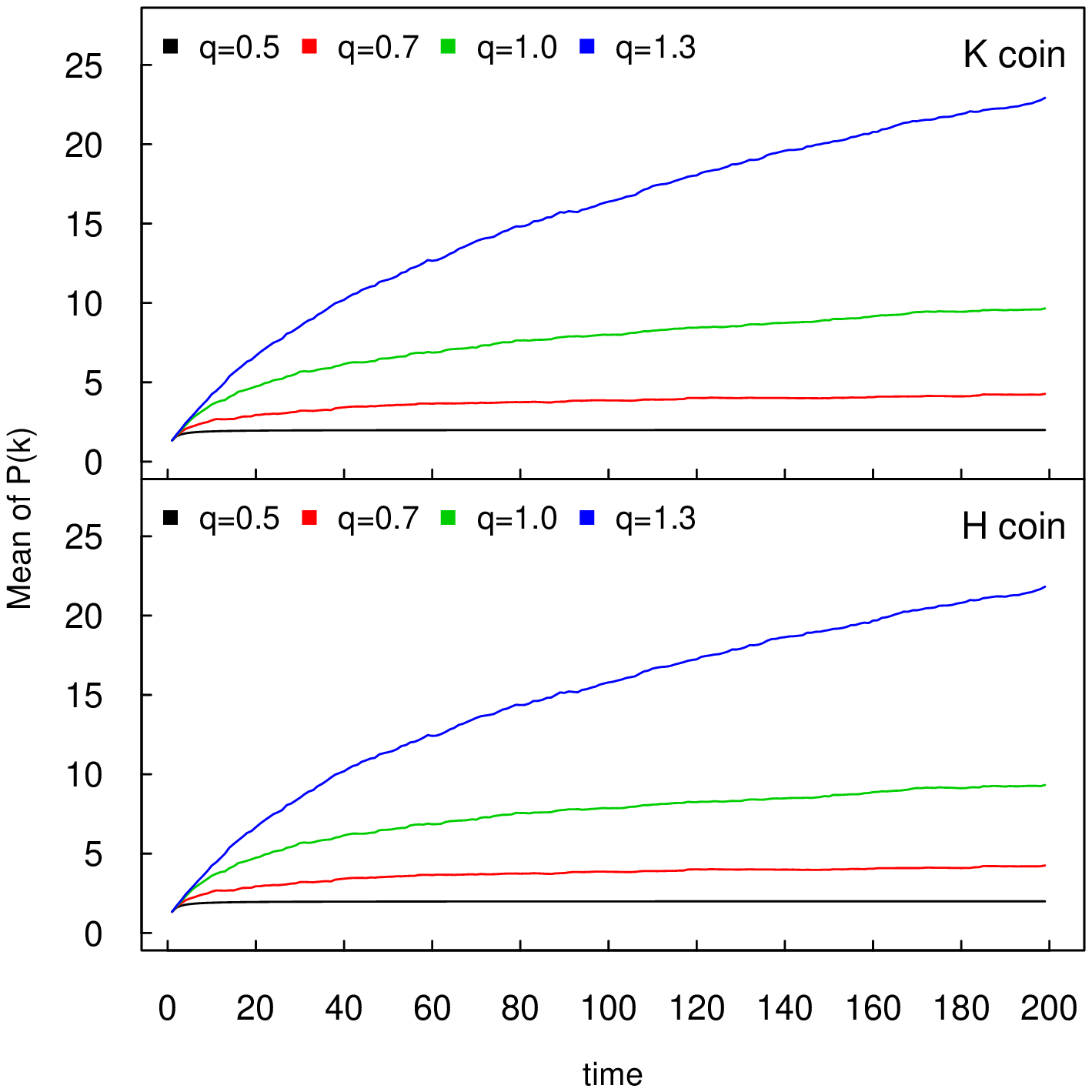}
\includegraphics[scale=0.48]{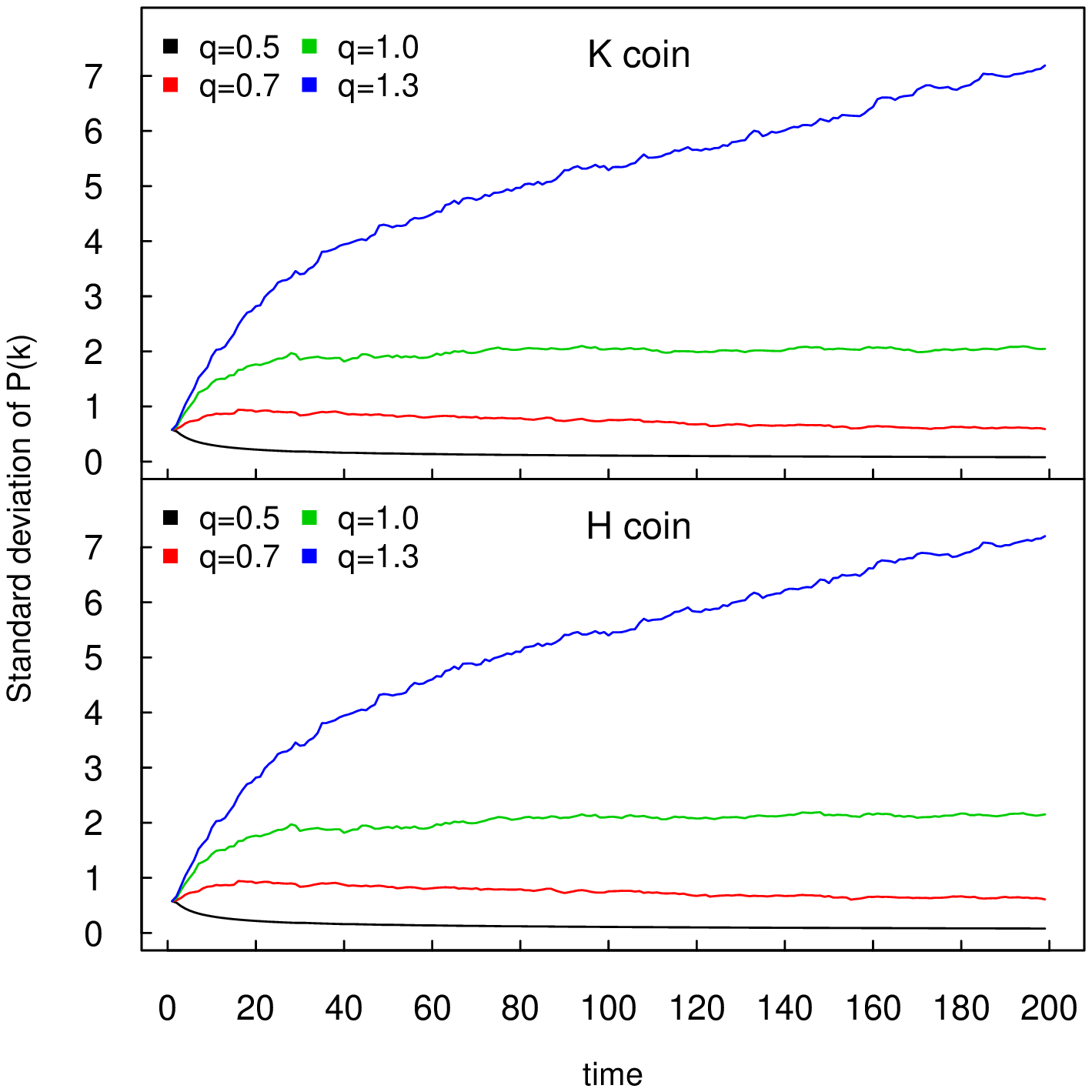}

\includegraphics[scale=0.48]{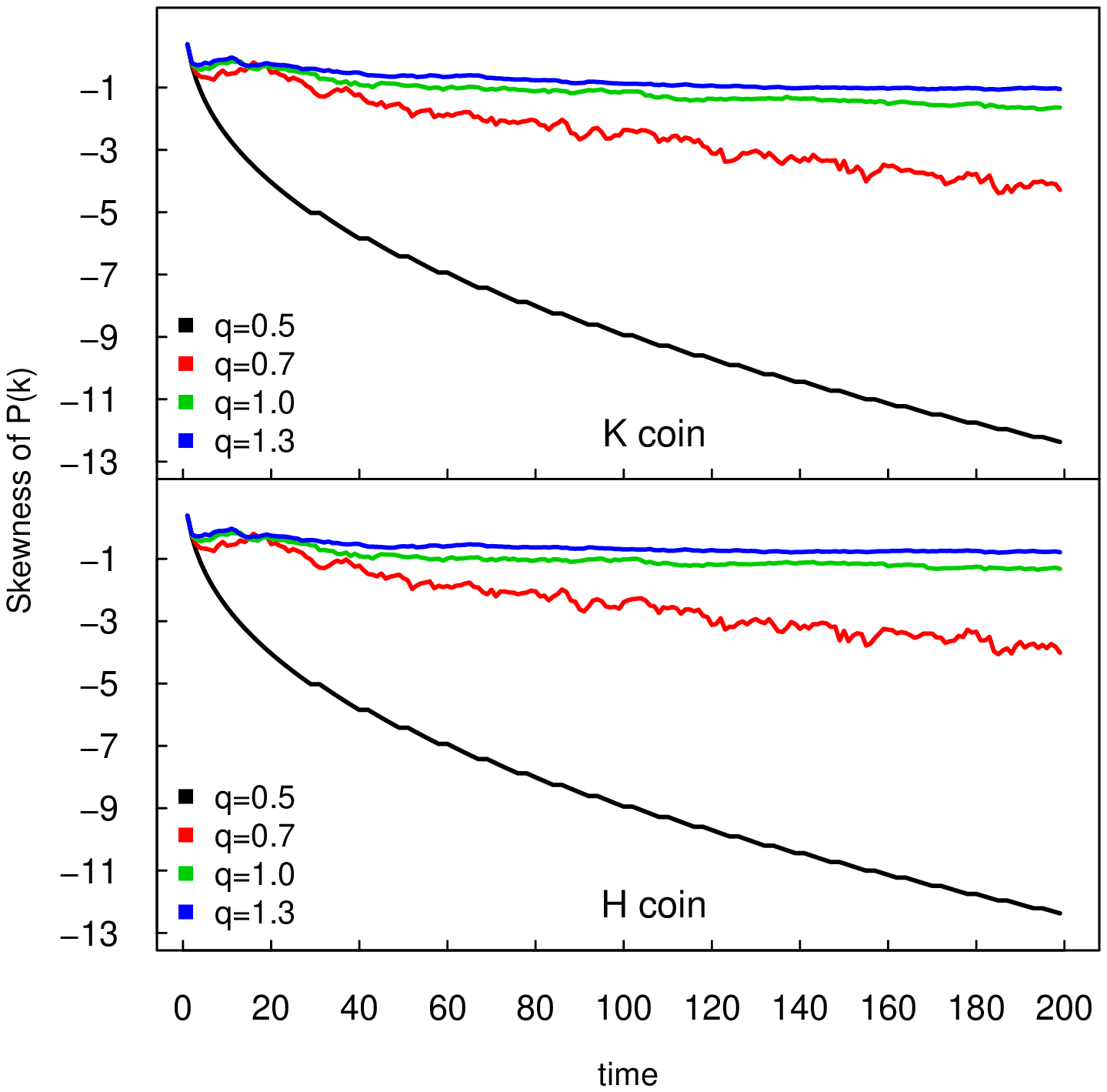}
\includegraphics[scale=0.48]{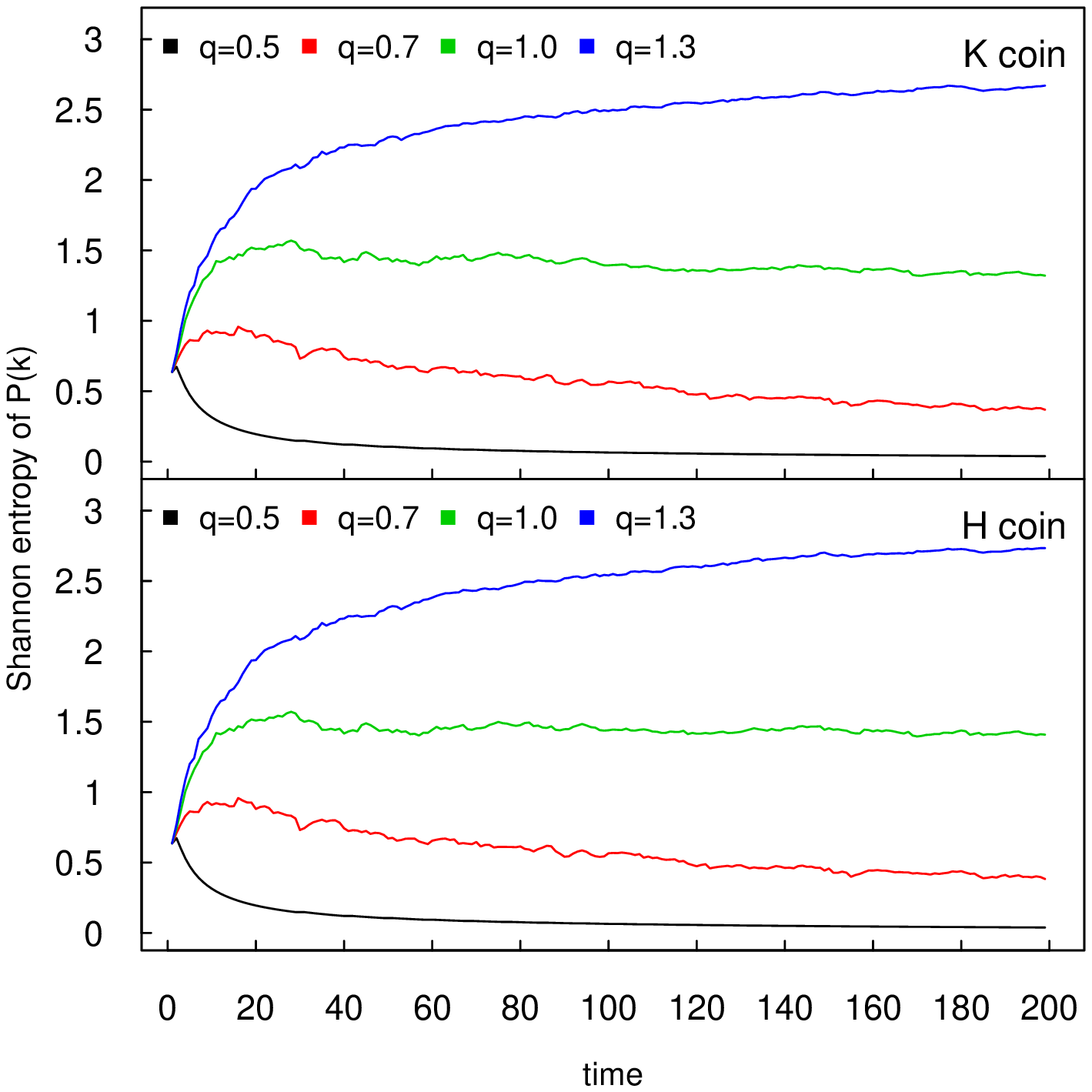}

\caption{Time series of the statistical properties of the degree distribution P(k). Panels show typical time-evolutions for mean, standard deviation, skewness and Shannon entropy of P(k).}
\label{fig:net-statistical}
\end{figure}

\begin{figure}[t]
\centering
\includegraphics[scale=0.48]{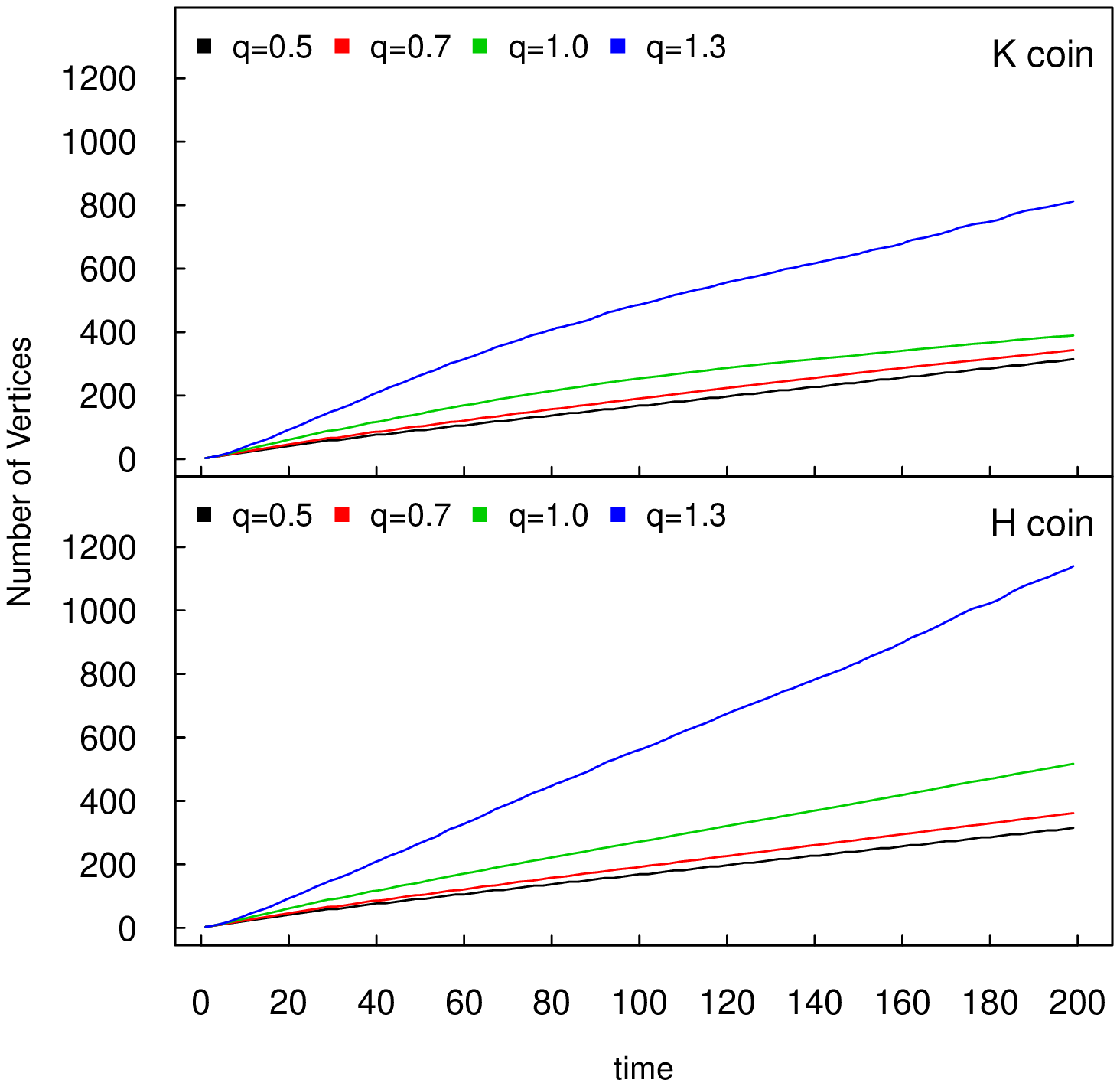}
\includegraphics[scale=0.48]{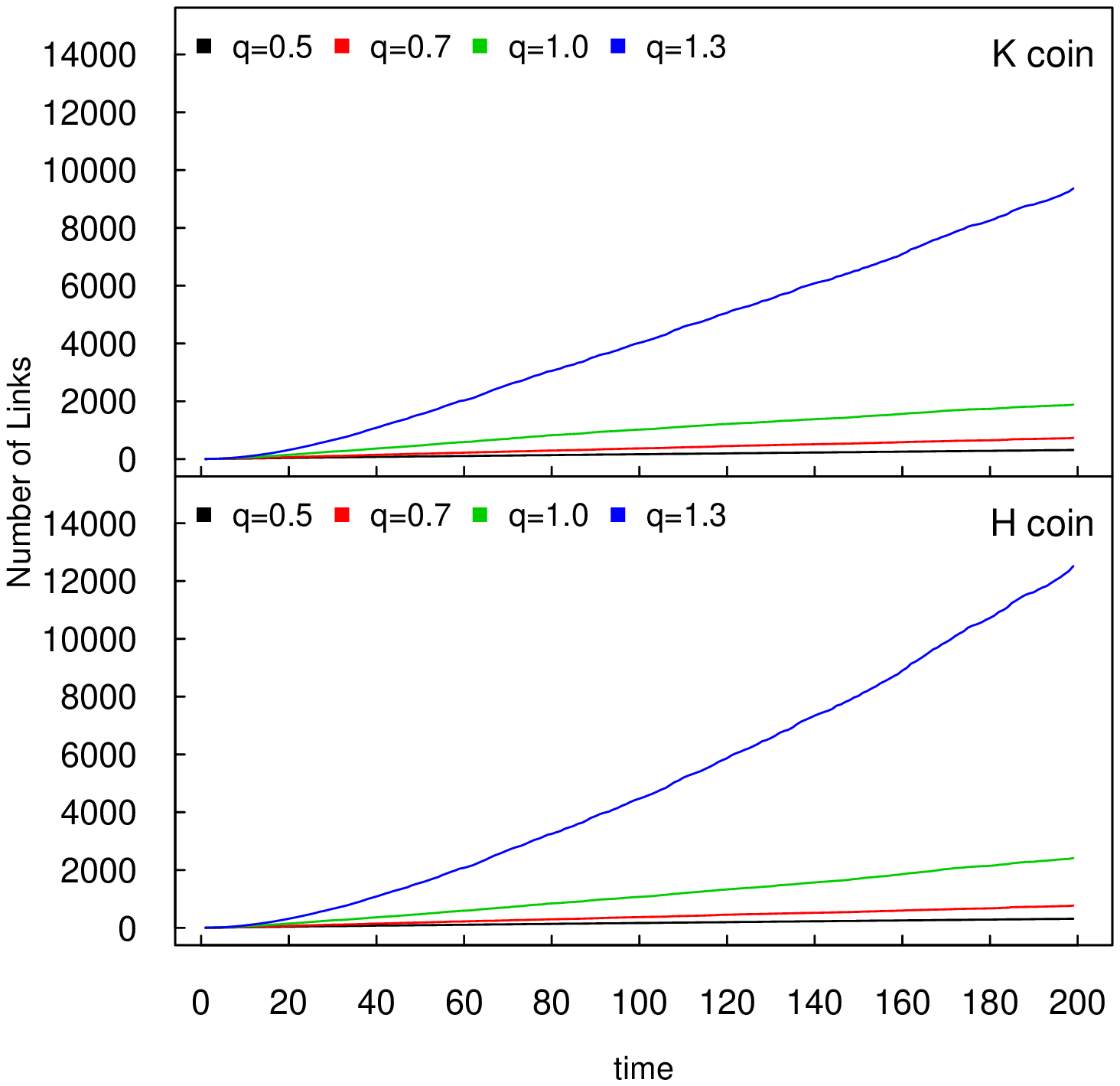}

\includegraphics[scale=0.48]{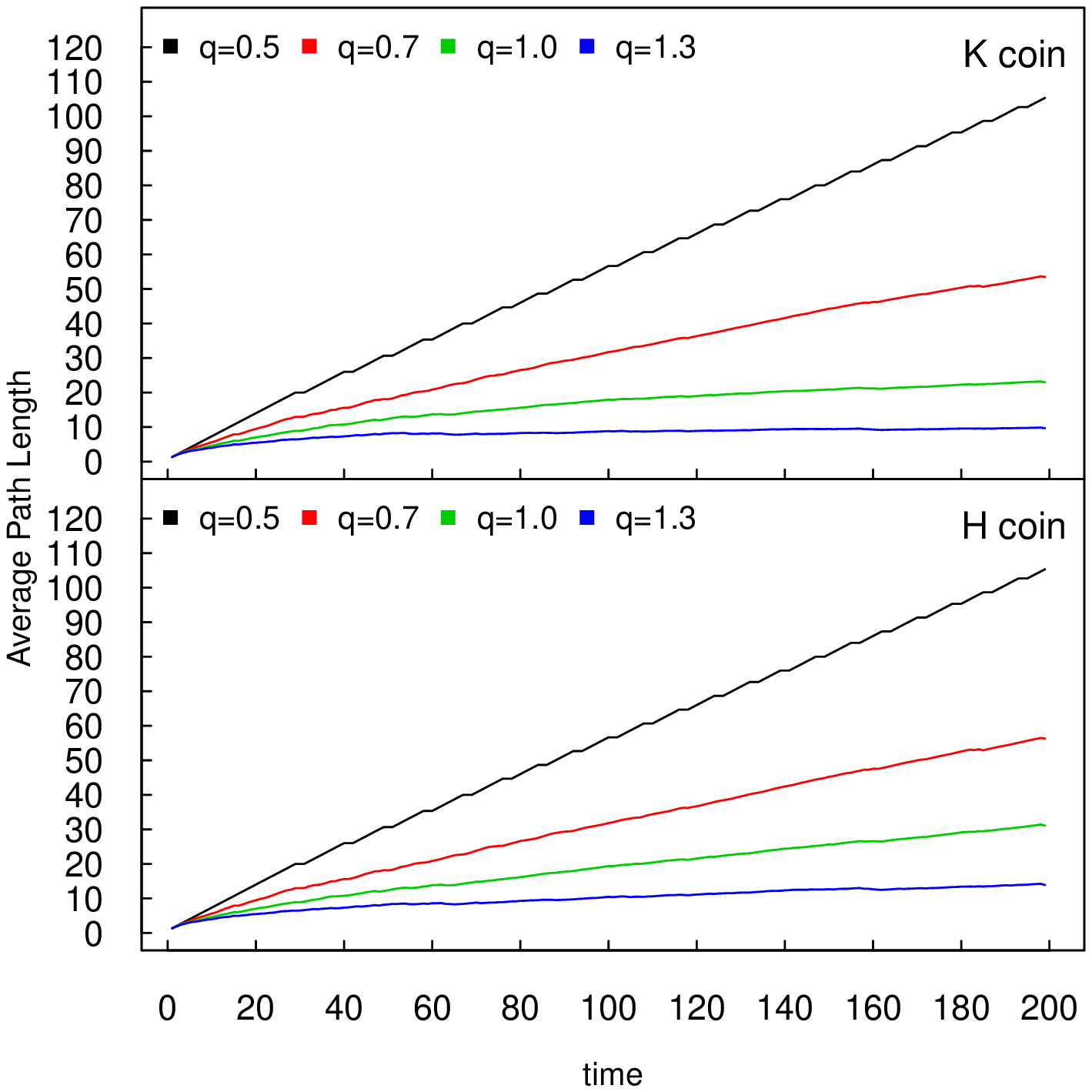}
\includegraphics[scale=0.48]{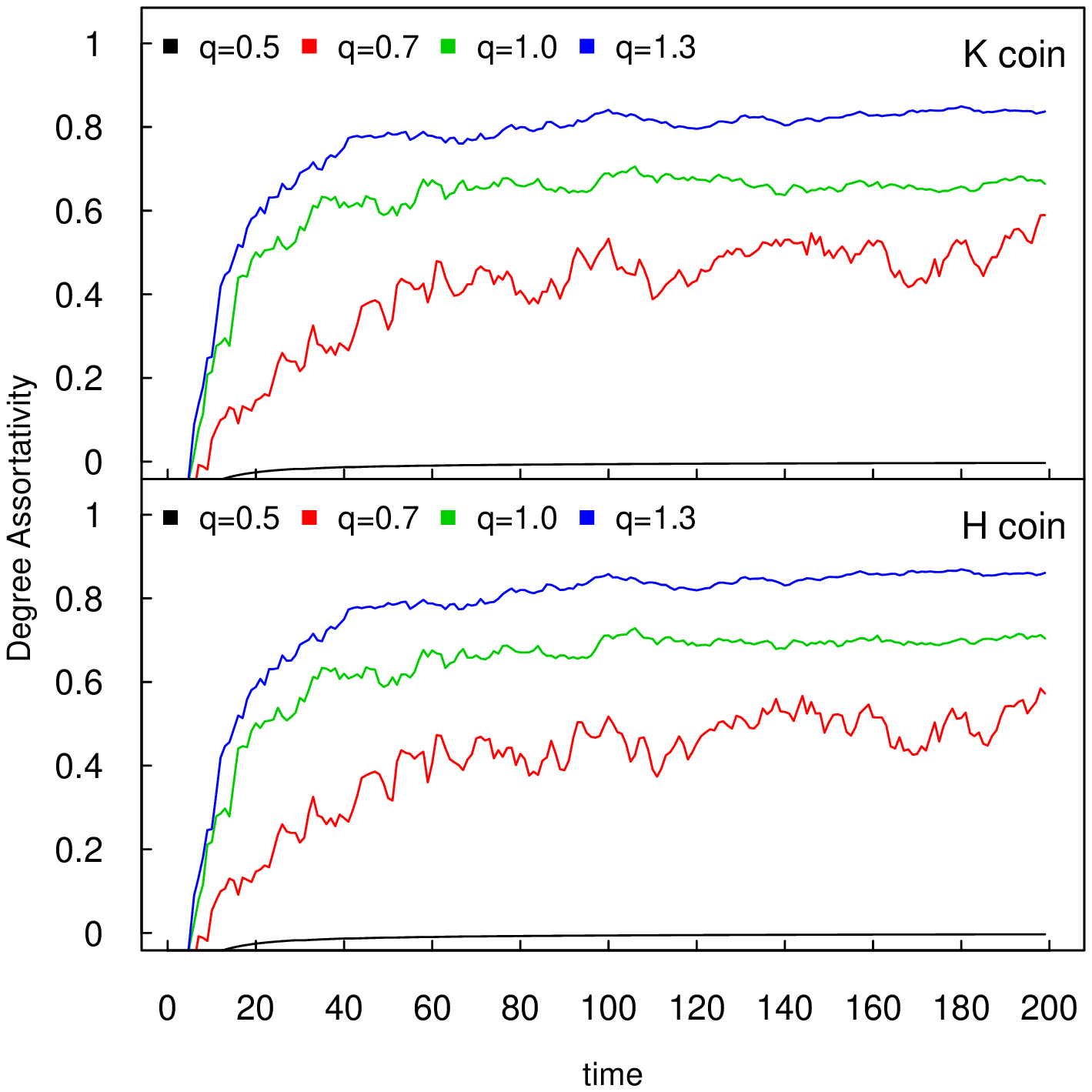}

\caption{Time series of the graph-based properties. Panels show typical time-evolutions for number of vertices and links as well as average path length and degree assortativity.}
\label{fig:net-structural}
\end{figure}

We start by presenting our results for the diffusion of the packet that is computed from the second statistical moment at time $t$,
\begin{equation}
\overline{x^2} _t =\sum_x x^2P_t(x) 
\end{equation}
where
 \begin{equation}
 P_t(x) = |\psi_{t}^{L}(x)|^2 + |\psi_{t}^{R}(x)|^2.
 \end{equation}
As we deal with initial conditions that yield symmetric $ P_t(x)$, it follows that $\overline{x^2}$ is indeed the variance 
$\sigma^2 = \overline{x^2} - \overline{x}^2$  since $\overline{x}=0$.

 To evaluate the  dynamical regimes we have computed the scaling exponent $\alpha$ of the power-law $\overline{x^2}\propto t^\alpha$, with $t\gg 1$. In Fig.\ref{fig:diagram-alpha}, we see that the gEQW exhibits a  rich dynamics ranging from diffusive $ \alpha =1$  to hyperballistic behaviour $ \alpha = 3 $ for uniform(long-range) memory, which is approached as we increase the value of $q$, especially for $q \gtrapprox 4/3$.
 On the other hand, it is visible the standard QW can be understood a particular case in the relation between the functional form of the memory and the sort of diffusion performed by the walker because it corresponds to a singularity. 
 In respect of that, we have performed a battery of tests in the vicinity of $q$ that confirmed our stance.
We see that the memory acts a double-edged sword: for $ 1/2 < q \lessapprox 4/3 $, the diffusion exponent decreases for $C_H$ and remains largely diffusive using $C_K$, whereas for $q \gtrapprox 4/3 $ one has an augment of $\alpha $; moreover, the difference between the two coins vanishes from that value of $q$ on.  

What is the reason for this twofold non-increasing/increasing behaviour of the diffusion exponent with $q$? At the fundamental level, we assert that it stems from the quantum superposition of the states during the evolution of the gEQW. We base our reasoning by computing the Relative Quadratic Deviation
\begin{equation}
RQD(x) \equiv  (x - \bar{x} )^2 P_t(x),
\label{eq:rqdx_1}
\end{equation}
where $\bar{x}=\sum_x x \, P_t(x) $.

We regard $RQD(x)$ as a local measure gauging the contribution of each site to the global variance. The results of $RQD(x)$ and $P_t(x)$ for different values of $q$ are plotted in Fig.~\ref{fig:pxt}-\ref{fig:rwdx} and interpreted as follows:
\begin{itemize}
\item for $q=1/2$ --- the memoryless case ---, all the steps have the same size equal to $1$. The Quantum Walker does not spread as much as the cases for $q > 1/2$, which leads to relatively small values of $(x - \bar{x} )^2$. However, the  distribution $P_t(x)$ exhibits peaks concentrated near the borders. This combination generates a $RQD(x)$ profile with some large values, which are the key contributors to the ballistic spreading. 

\item For $q=1$ --- the exponential case---, we note that $(x - \bar{x} )^2$ achieves higher values than the previous case because it is possible to have large steps that increase the occupation of positions $x$ far away from the origin. Inasmuch as $(x - \bar{x} )^2$ achieves larger values, $P_t(x)$  handicaps the sites near the borders giving rise to relatively small values of $RQD(x)$. Conversely, the peaks of the probability near the origin are weakened by $(x - \bar{x} )^2$ leading to  small values $RQD(x)$ as well;

\item For $q=1.5$ --- in the asymptotic power-law regime --- the wavepacket occupies substantially more sites; however, the site occupation is much less localised than in the preceding cases. The sites far off from the origin naturally have values of $(x - \bar{x} )^2$ that are large enough to overcome its little probability $P_t(x)$. This sets forth a profile for $RQD(x)$ that contains very large values, which induce the boosting of the gEQW. 

\end{itemize} 

To characterise the localisation of the wavefunction we have used two complementary measures, namely the Shannon entropy $S$
\begin{equation}
S =  - \sum_x   P_t(x)  \log  P_t(x),
\end{equation}
and the Inverse participation ratio (IPR)
\begin{equation}
IPR = \frac{1}{\sum_x \left[P_t(x) \right]^2},
\end{equation}
which quantifies the typical number of lattice sites that participate in wavepacket.

We applied both measures because such approach allows a more comprehensive assessment of the localisation in the model. In other words, the extreme cases of both measures have well-defined values:
(i) fully localized states $ \rightarrow P_t(x)=\delta_{x,0} \rightarrow S=0$ and $IPR=1$;
(i) fully delocalized states$ \rightarrow P_t(x)=1/N \rightarrow S=\log N$ and $IPR=N$. However, the two measures can behave differently from one another. Due to its quadratic term, the $IPR$ is prone to overvalue the sites with dominant $P_t(x)$ whereas the entropy $S$ --- owing to its logarithmic dependence --- gives relevance to sites with little probability $P_t(x)$. 

As visible in Figs.~\ref{fig:IPR}~and~\ref{fig:occupancy}, an increase in the memory exponent $q$ generates an overall increase in $S$, $IPR$; therein, we depict the Occupancy\cite{dan2015one} of the lattice, 
\begin{equation}
Occ= \sum _x \Theta \left( P_t(x) \right),
\end{equation}
i.e., the number of sites with non-zero probability (numerically $>10^{-9}$). In Fig.\ref{fig:occupancy} the overall pattern of  $Occ$  with $q$ is increasing as well, however $Occ$ displays a more pronounced difference for the two coins used. This is caused by the presence of  non-negligible local maxima off the center of the grid in the H coin case, as illustrated in~Fig\ref{fig:pxt}.

In Fig.~\ref{fig:rwdx} we see that jumps give rise to a dissimilarity between the probability distributions obtained by utilising  either the $H$ or the $K$ coin. In order to assess how this effect evolves, we have computed the Jensen-Shannon dissimilarity (JSD)
\begin{equation}
JSD = \frac{KLD(P^{h}|M) +  KLD(P^{k}|M) }{2}
\end{equation}
where $M(x)=(P^{h}(x) + P^{k}(x) )/2$ corresponds to the  point-wise midpoint  distribution  between the probability distributions $P^{h}(x)$ and $P^{k}(x)$. The JSD  is based on a simmetrisation of the Kullback-Leibler dissimilarity (KLD) between two distributions $U$ and $W$
\begin{equation}
KLD(U|W)= \sum_{x} U_x \log _{2} U_x/W_x ,
\end{equation}
with the advantage that in the former the domain of  distributions can be different without yielding an incommensurable result.
In Fig\ref{fig:jsd}, we see that, as time elapses, the memoryless case ($q=0.5$)  is associated with similar distributions
$P_t^{Hcoin}(x)$  and $P_t^{Kcoin}(x)$ and thus $JSD=0 \ \forall t$. However, the introduction of memory changes that picture with noticeable sensibility of the probability distribution --- hence the diffusion --- to the coin operator.

In addition, we have evaluated the entanglement between the internal and external degrees of freedom 
(the so-called coin-space entanglement by means of the Von Neumann entropy
\begin{equation}
S_e  = - \mathrm{Tr} \left[ \rho^c\log\rho^c \right],
\end{equation}
where $\rho^c = \mathrm{Tr}_x(\rho) $ is the reduced density matrix of the particle and $\rho$ is the full density matrix $\rho = | \Psi \rangle \langle  \Psi | $ of the QW system. An explicit expression was obtained previously (for instance see~\cite{zeng2017discrete,abal2006quantum}).
\begin{equation}
\rho^c(t) = 
\begin{bmatrix}
G_a & G_{ab} \\
G^{*}_{ab} & G_b
\end{bmatrix},
\end{equation}
where $G_a = \sum_x |\psi_{t}^{L} (x)|^2 $, 
$G_b = \sum_x |\psi_{t}^{R} (x)|^2 $ and 
$G_{ab} = \sum_x \psi_{t}^{L}(x)  \left( \psi_{t}^{R}(x)  \right)^{*}   $.
The entropy $S_e$ is actually determined resorting to the eigenvalues  $\lambda^{\pm}$ of $\rho^c$     
\begin{equation}
 S_e  = -\lambda^{-} \log_2 \lambda^{-} - \lambda^{+} \log_2 \lambda^{+}
\end{equation}
\begin{equation}
 \lambda^{\pm} = \frac{1}{2} \pm \frac{1}{2}\sqrt{1-4G_aG_b + 4|G_{ab}|^2 }
\end{equation}

Interestingly, in Fig.~\ref{fig:entang} we can understand that for all $q>1/2$ one finds a very large entanglement entropy $ Se\approx 1 $ in the long-run. Similar results were obtained in Ref.~\cite{chandrashekar2012disorder,vieira2013dynamically} 
however, the authors assumed a  time-dependent disorder in the coin operator. 
But epistemologically, both model fits within the class of QWs with the temporal disorder.

\subsection{From quantum walks to complex networks}

As time elapses, the quantum walker visits sites on the grid so that we can map those displacements into a network. To do so, we start the network with one node corresponding to the central position $x=0$ and set $\mathcal{N}_i$ as the neighbour of the focal node $i$. In the growing process, we use nondirectional edges for the nodes $i, j$. The network is then governed by the algorithm:

\begin{itemize}
\item For each $t \geq t_0+1 $: apply the model described in Section \ref{sec:model} and subsequently: 
 \begin{itemize}
  \item For every position $i=x$ with  $P_{t}(i) > 0 $:
  \begin{itemize}
   \item For $j \in \{x-dx,x+dx\}$: 
   \begin{itemize}
     \item  If  $P_{t-1}(j) > 0 $ and $j \notin \mathcal{N}_i$: add $j$ to the neighbour list of the focal node $i$, ie create a link between the sites $i$ and $j$, since 
   \end{itemize}
  \end{itemize}
 \end{itemize}
\end{itemize}

This map QW $\rightarrow$ network allows using a powerful graph theory toolbox to better grasp the underlying phenomena behind the gEQW. In the current case, we have opted for  the R-Igraph  and Python-Networkx packages.
In Fig.~\ref{fig:net-degree}, we see the degree distribution, $P(k)$, of the typical networks emerging from different  gEQW  tend to be very unequal and left-skewed for small $q$ and more uniformly distributed for high $q$. This pattern is well captured by both the Shannon entropy and the skewness of $P(k)$ where the first increases as the memory range parameter $q$ increases and the second become less negative over time (see Fig.\ref{fig:net-statistical}). We have seen the mean degree and standard deviation of $P(k)$ increases with $q$, because of small $q$ act as an effective barrier for long-range hopping events.

In Fig.~\ref{fig:net-structural}, we show how the main structural properties of the QW network evolve over time for increasing $q$. Both the number of links $n_L$ and vertices $n_V$ increases, with different concavity though. On the other hand, the average path length  exhibits an overall decreasing pattern with $q$ since long-range jumps tend to establish links between two nodes far away from each other.

Among all the measures, the degree of assortativity experiences a sharp rise in the short-run  and then it grows more slowly.  The overall positive value means  that the degree of the vertices correlates across the network. For moderated jump range such as for $q=0.7$, the fluctuations tend to be very high in comparison with the other cases. For comparison remember that random networks have assortativity degree  near zero, meaning the absence of degree-degree correlations across the whole network.

Curiously, the overall outcomes for $H$ and $K$ coins in Fig.\ref{fig:net-statistical}-Fig.\ref{fig:net-structural} are much less affected by the coin operator. This is a remarkable difference from the results presented in the previous section, which were much more impacted by the coin operator (see Fig\ref{fig:diagram-alpha} for small q). In that sense, the graph approach provides a different perspective to investigate QWs with jumps.

\section{\label{sec:Conclusion}Concluding Remarks}

In this paper, we have comprehensively explored an extension of the EQW, a non-Markovian quantum walk process where the jump sites are selected from a uniform distribution giving rise to hyperballistic diffusion. Discrete-time QW with short-range jumps has been addressed in 
\cite{lavivcka2011quantum,das2019inhibition,sen2019scaling,sen2019unusual}. The authors found that jumps of small size lead to an inhibition of spreading as we also have found for small $q$. But none of these previous works have found the richness of dynamical transitions we have presented herein with our new time-dependent protocol of jumps and two types of coin operators. Our flexible distribution, the $q$-exponential,  allows us to recover both the standard QW in the limit $q \rightarrow 1/2$ and the previous proposal $q \rightarrow \infty$\cite{di2018elephant}, providing a more general framework.

Since the model is about a walker, we first focussed on its diffusion properties. We have verified that the functional form of the distribution from which the jumps are drawn impacts in the kind of diffusion exhibited by the model. Particularly, for $q^* \lesssim 4/3 $ the walker is sub-ballistic with specific behaviour depending on the sort of quantum coin that is applied. Concretely, setting apart the case $q = 1/2$ for which the standard QW is retrieved whatever the case, for the Hadamard-type coin the diffusion slowly decays from an exponent value slightly larger than the superdiffusion case $\alpha = 1.5$ and from $q^*$ onwards the dynamics changes to a consistent increase of the diffusion exponent with $q$ until it reaches the hyperballistic regime when $q \rightarrow \infty$. On the other hand, the non-hermitian coin yields a normal diffusive regime --- within statistical significance --- for $q < q^*$ and therefrom the value
diffusion exponent soars up to the EQW case. 
We assign to the eigenvalue structure introduced by $C_H$ and $C_K$ the difference in the diffusion behaviour that is developed in each case.

Although we did not manage to find an analytical evidence thereof, we cannot help mentioning the values $q$ whereat the model both starts to augment $\alpha $ and recovers ballistic diffusion (with $q \neq 1/2$) correspond to the cases in which $q$-exponential distribution stops having finite standard deviation and average, respectively.
Nevertheless, resorting to a simple measure derived from the variance we have been able to learn dependence between the memory distribution and the diffusion exponent. With respect to information related measures, we have verified that the asymmetry of the entropy with the angle of the coin clearer; moreover, the larger $q$ the larger the entropy for small angles. An equivalent qualitative relation is found for quantities such as the inverse participation ratio and the occupancy.

Concerning the entanglement produced in the model, the analysis of the density matrix eigenvalues we have verified a very large entanglement for all of the values of $q> 1/2$. Similar results were obtained in systems assuming disorder in the coin operator. Thus, we reckon that strong entanglement is a feature of QWs with some sort of randomness.
At first, it seems counterintuitive that by introducing disorder it is possible to establish quite entangled states; however, it is actually this randomness that creates the change of overlap between the states so that the density matrix does not show a pure state structure. If we understand the emergence of entanglement as a manifestation of some kind of order --- that can be in the form of propagation of information ---, our results help characterise such class of models the quintessential complex system. In other words, we have microscopic details yielding a robust (and unexpected) macroscopic feature. Owing to its jump structure between different possibilities, this QW could be viewed from a network perspective. Carrying out a network analysis we have computed that the jump network broadens significantly as the value of $q$ gets larger but we have no evidence that the degree distribution exhibits fat tails.

\begin{acknowledgments}

The authors acknowledge financial support from the Brazilian funding agencies CAPES (MAP) as well as CNPq and FAPERJ (SMDQ). GDM also wishes to thank the fruitful discussions with Armando C. Perez and Alberto Verga.

\end{acknowledgments}

\bibliography{apssamp}

\end{document}